\documentclass[runningheads]{llncs}
\def\BibTeX{{\rm B\kern-.05em{\sc i\kern-.025em b}\kern-.08em
    T\kern-.1667em\lower.7ex\hbox{E}\kern-.125emX}}
\usepackage{graphicx}
\usepackage{mathtools}
\usepackage{lipsum} 
\usepackage[numbers]{natbib}

\usepackage{booktabs}
\usepackage[ruled]{algorithm2e}
\usepackage{url}
\usepackage{color}
\usepackage{multirow}
\usepackage{enumitem}
\usepackage{array}
\usepackage{siunitx}
\usepackage[labelfont={bf}]{caption}
\usepackage{balance}
\usepackage{array} 
\usepackage{todonotes}
\usepackage{subfig}
\usepackage[nolist,nohyperlinks]{acronym}
\usepackage{footnote}
\usepackage[perpage]{footmisc}
\usepackage[english]{babel}
\usepackage[autostyle]{csquotes}
\makesavenoteenv{tabular}

\acrodef{CNN}{Convolutional Neural Network}
\acrodef{AI}{Artificial Intelligence} 
\acrodef{ML}{Machine Learning}
\acrodef{DL}{Deep Learning}
\acrodef{DNN}{Deep Neural Network}
\acrodef{DDANet}{Dual Decoder Attention Network}
\acrodef{CADx}{Computer Aided Diagnosis}
\acrodef{GI}{Gastrointestinal}
\acrodef{DSC}{Dice Coefficient}
\acrodef{mIoU}{mean Intersection over Union}
\acrodef{FPS}{Frame per Second}
\begin{document}

\title{DDANet: Dual Decoder Attention Network for Automatic Polyp Segmentation}

\titlerunning{DDANet}
\author{Nikhil Kumar Tomar \inst{1} \and
Debesh Jha\inst{1,3}\and
Sharib Ali \inst{4}\and
H{\aa}vard D. Johansen\inst{3} \and
Dag Johansen\inst{3} \and
Michael A. Riegler\inst{1} \and
P{\aa}l Halvorsen\inst{1,2}
}

\authorrunning{Tomar et al.}

\institute{
    SimulaMet, Norway \and
    Oslo Metropolitan University, Norway \and
    UIT The Arctic University of Norway \and
    Department of Engineering Science, University of Oxford, Oxford, UK\\
    \email{debesh@simula.no}}
\maketitle            
\begin{abstract}
Colonoscopy is the gold standard for examination and detection of colorectal polyps. Localization and delineation of polyps can play a vital role in treatment (e.g., surgical planning) and prognostic decision making. Polyp segmentation can provide detailed boundary information for clinical analysis. Convolutional neural networks have improved the performance in colonoscopy. However, polyps usually possess various challenges, such as intra-and inter-class variation and noise.  While manual labeling for polyp assessment requires time from experts and is prone to human error (e.g., missed lesions), an automated, accurate, and fast segmentation can improve the quality of delineated lesion boundaries and reduce missed rate. The Endotect challenge provides an opportunity to benchmark computer vision methods by training on the publicly available Hyperkvasir and testing on a separate unseen dataset. In this paper, we propose a novel architecture called ``DDANet'' based on a dual decoder attention network. Our experiments demonstrate that the model trained on the Kvasir-SEG dataset and tested on an unseen dataset achieves a dice coefficient of 0.7874, mIoU of 0.7010, recall of 0.7987, and a precision of 0.8577, demonstrating the generalization ability of our model.  


\end{abstract}
\vspace{-5mm}
\section{Introduction}
\label{sec:introduction}
\vspace{-3mm}
Colorectal cancer is one of the leading causes of cancer. Colonoscopy is a standard medical procedure for the surveillance examination and treatment. Regular screening and removal of pre-cancerous lesions through colonoscopy is essential for early cancer detection and prevention. Studies suggest that the miss-rate of adenoma is between  6 to 27\%~\cite{ahn2012miss}.

The automatic segmentation of the suspected areas with lesions in colonoscopy images can play a crucial role, and identifying each colon pixel can significantly impact clinical settings. With the increase of publicly available datasets, dominant methodology such as convolutional neural network, improved hardware, and  collaboration between computational and clinical communities to tackle the problems in endoscopic imaging through computer vision tasks is gaining momentum than ever before.  An automatic polyp detection or surveillance system can help to achieve low-cost design solutions and save time of clinicians allowing them to use their time to look into more severe cases.

In this respect, the Endotect challenge~\cite{Hicks2020} offers three tasks, namely, detection of \ac{GI} tract images, efficient detection on the same images,  and automatic polyp segmentation. The detection and efficient detection task are based on the HyperKvasir dataset~\cite{borgli2019hyper}, and the segmentation is based on the Kvasir-SEG dataset~\cite{jha2020kvasir}. Out of these three tasks, we participated in the ``segmentation task", where the goal was to generate an automatic segmentation of the polyps for the unseen dataset. 

In this paper, we propose a novel deep learning architecture, called \ac{DDANet}, for automatic polyp segmentation. It follows an encoder-decoder scheme and incorporates a single encoder that is shared by two parallel decoders, where the first decoder acts as a segmentation network and the second decoder acts as an autoencoder network. The autoencoder network helps to strengthen the feature maps in the encoder network. It is used as an auxiliary task training, which is used to generate an attention map. This attention map is used in each decoder to improve the semantic representation of the feature maps. This, in turn, helps to improve the performance of the entire network. The proposed \ac{DDANet} is fed with an RGB input image,where it predicts the segmentation mask and the reconstructed grayscale image. The architecture is efficient in terms of \ac{FPS} and also has a decent evaluation score. These metrics are the requirement for the real-world settings toward developing a \ac{CADx} system.

\section{Related Work}
\label{sec:relatedwork}
\vspace{-3mm}
Automatic polyp segmentation task is a well-defined computer vision problem. Recently, there have been several competitions~\cite{bernal2017comparative,ali2020objective,ali2020translational,jha2020medico} and individual efforts~\cite{guo2020polyp,jha2019resunet++,Jhadoubleunet2020,guo2019giana,jharealtime} toward building a \ac{CADx} system for the polyp segmentation. With these competitions and individual efforts, polyp segmentation is becoming more and more mature. However, comparing models and results of the many individual approaches is difficult due to the use of diverse (often publicly non-available) datasets and different hardware. In this respect, competitions provide an opportunity to benchmark and compare the designed methods with other competitors' on the same dataset. Moreover, the evaluation metrics are independently calculated by the organizers, including the ranking decision of each team.

The competitions can help us to define the strengths and weaknesses of each method. It also provides us with an opportunity to disseminate methods and discuss the results collectively in the same space. Through this year's Endotect challenge, we provide a novel solution to develop more efficient algorithms that can be useful to build an automatic polyp segmentation system. 
Our architecture is composed of an autoencoder branch in addition to the segmentation branch, which is different from other encoder-decoder based network (for example, UNet~\cite{ronneberger2015u}, ResUNet++~\cite{jha2019resunet++}, DoubleUNet~\cite{Jhadoubleunet2020}). The benefit of incorporating autoencoder in the network can be seen from the quantitative and qualitative results. 
\begin{figure}[t!]
    \centering
    \includegraphics[width = 1.0\columnwidth]{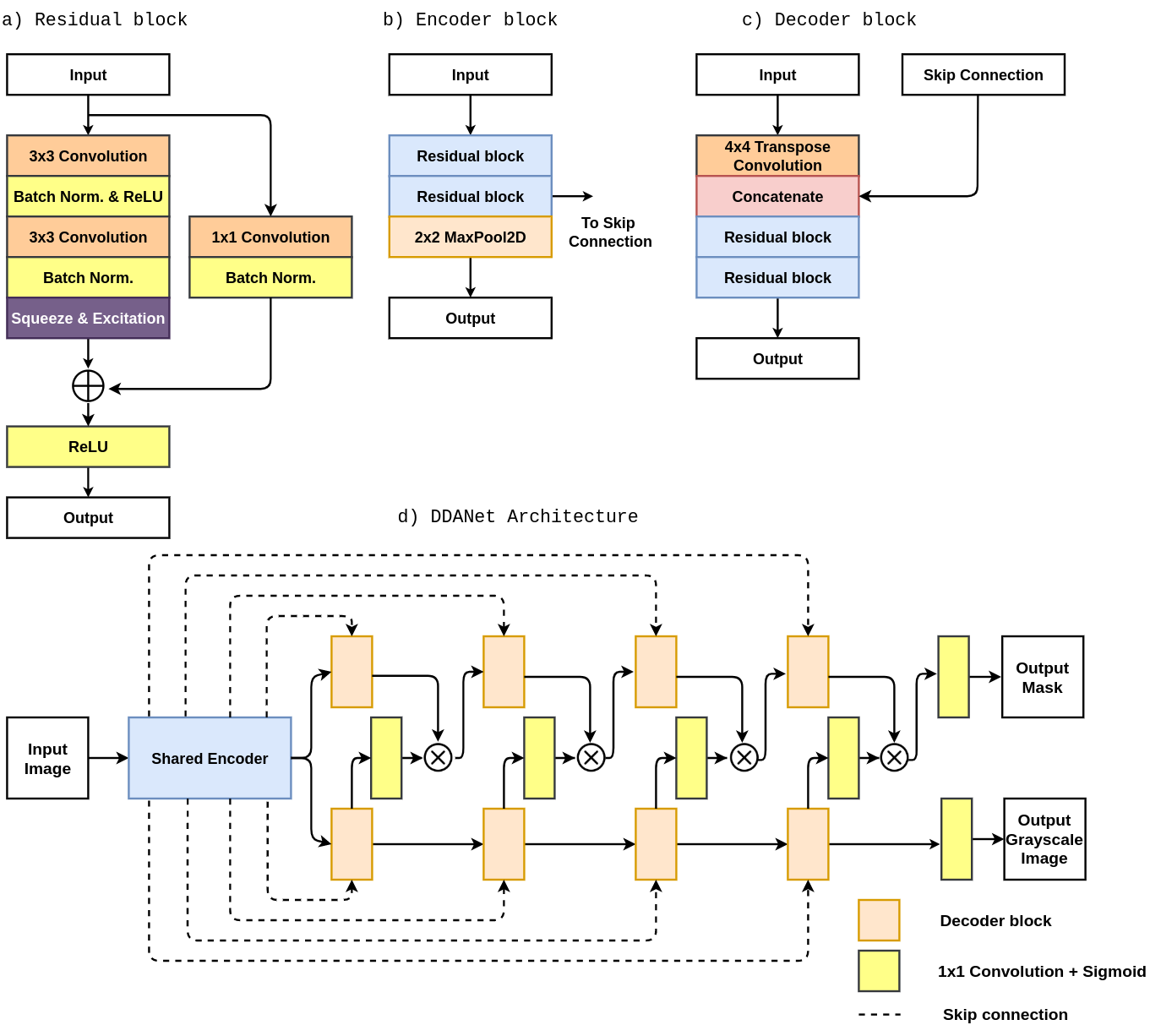}
    \caption{DDANet architecture and its components.}
    \label{fig:my_label}
\end{figure}
\section{DDANet}
\label{sec:Methodology}
In this section, we will first describe each component of our DDANet and then detail the overall proposed DDANet architecture.
\subsection{Residual block}
As the network depth increases, the performance also increases to a certain limit as the gradients can be effectively calculated. However, after a certain depth, the performance of the model may be impacted due to the vanishing or exploding gradients as the gradients become either zero or too large. By introducing a skip-connection in residual learning, the problem of the vanishing or exploding gradients has been solved. Our residual block (see Figure~\ref{fig:my_label}a) consists of two $3\times3$ convolutions, each followed by a batch normalization and a Rectified linear unit (ReLU) activation function. 
The residual learning introduces a shortcut connection or identity mapping, which connects the input with the residual block's output. The identity mapping tries to learn an identity function since the input is directly passed to the output. It also helps in a better flow of the gradients during the backpropagation.
\subsection{Squeeze and Excitation block}
\vspace{-2mm}
A \ac{CNN} is used to extract features from an image and then transform the image into a feature map. A problem with \acp{CNN} is that they treat every feature channel as equally important. To overcome this problem, we introduce a squeeze and excitation layer, which acts as a channel-wise attention mechanism. It re-weights every feature channel accordingly to create a more accurate feature map. In this way, the overall network becomes more sensitive towards essential features that improve the network performance significantly. The squeeze and excitation network mainly consists of two steps. In the first step, the feature maps are compressed using the global average pooling function to generate a compressed representation for the feature maps. While, in the second step, a $2$-layered neural network is used, where features are first reduced and then expanded. This generates a feature vector, which is used to scale the feature channels. 
\subsection{The DDANet architecture}
\vspace{-2mm}
The proposed architecture named \ac{DDANet} follows an encoder-decoder design similar to ResUNet++~\cite{jha2019resunet++}. The \ac{DDANet} combines the strength of the residual learning and the squeeze and excitation network. The proposed \ac{DDANet} is a fully convolutional network that consists of a single encoder shared by dual decoders. The encoder network consists of a $4$ encoder block, whereas each decoder network also consists of $4$ decoder block (see Figure~\ref{fig:my_label}d). 

The RGB input image is first fed into the encoder network (see Figure~\ref{fig:my_label}b), which encodes it into an abstract feature representation while gradually downsampling it. The output of the encoder network is fed to both decoders (see Figure~\ref{fig:my_label}c), where it is followed by a $4 \times 4$ transpose convolution that doubles its spatial dimensions. After that, the image is concatenated with an appropriate feature maps from the encoder network using the skip connection. These skip connections fetch the features from earlier layers at their original resolution, which increases their feature representation strength. The skip connections also act as an alternative path for the gradient flow and are often beneficial for model convergence. 

Two residual blocks are then used to learn the necessary feature required by the network during back-propagation. The output of the second decoder block (autoencoder branch) follows a $1 \times 1$ convolution and a sigmoid activation function to generates an attention map. This attention map is multiplied by the output of the first decoder block (segmentation branch), which acts as an input for the next decoder block in the segmentation branch. The final decoder block's output is passed through a $1 \times 1$ convolution and a sigmoid activation function, where the first decoder outputs a segmentation mask, and the second outputs the reconstructed grayscale image.

\section{Experimental Setup}
\label{sec:Experiments}
In this section, we present the implementation details and datasets used in this work. 
\vspace{-3mm}
\subsection{Implementation Details}
\vspace{-2mm}
The proposed \ac{DDANet} architecture is implemented in the PyTorch $1.6$ framework\footnote{\url{https://github.com/nikhilroxtomar/DDANet}}. For training the DDANet, we used an NVIDIA DGX-2 machine that uses an Nvidia V100 Tensor Core GPUs.During training, we have used an input image resolution of $512\times 512$. We use a combination of binary cross-entropy and dice loss for calculating the loss between the predicted masks and the ground-truth masks. We have used binary cross-entropy in the case of predicting the grayscale image. An Adam optimizer was used with a learning rate of 1e${^{-4}}$. The models were trained for 200 epochs. 

\subsection{Datasets}
\vspace{-2mm}
The Kvasir-SEG~\cite{jha2020kvasir} dataset was used for training the model. We have used 88\% of the dataset for training and the remaining 12\% images for development-test-set. Kvasir-SEG consists of 1000 polyp images, ground truth segmentation masks, and bounding boxes. A separate test dataset with 200 images was provided for prediction. However, the ground truth for this dataset was not provided by the organizers. The exact number of images used for the training and testing can also be found in our GitHub repository. More details about the dataset and the baseline results on it can be found in~\cite{jha2020kvasir}. 
\section{Results}
\vspace{-3mm}
\label{sec:Results}
 \begin{table}[t]
 \caption{Quantitative results on Kvasir-SEG and unseen (Challenge) dataset.}
    \label{table:resultkvasir}
    \def\arraystretch{1.5}
     \setlength\tabcolsep{3pt}
    \par\bigskip
    \centering
  \begin{tabular}{l|l|l|l|l|l|l}
\hline
\textbf{Dataset}&\textbf{Method} & \textbf{DSC} & \textbf{mIOU} & \textbf{Recall} & \textbf{Precision} & \textbf{FPS} \\ \hline
Kvasir-SEG& DDANet & 0.8576 & 0.7800 & 0.8880 & 0.8643 & 69.59\\ \hline
Unseen (\textbf{Challenge}) & DDANet & 0.7874 & 0.7010 & 0.7987 & 0.8577 & 70.23 \\ \hline

\end{tabular}
\end{table} 

\begin{figure}[t!]
    \centering
    \includegraphics[width = 1.0\columnwidth]{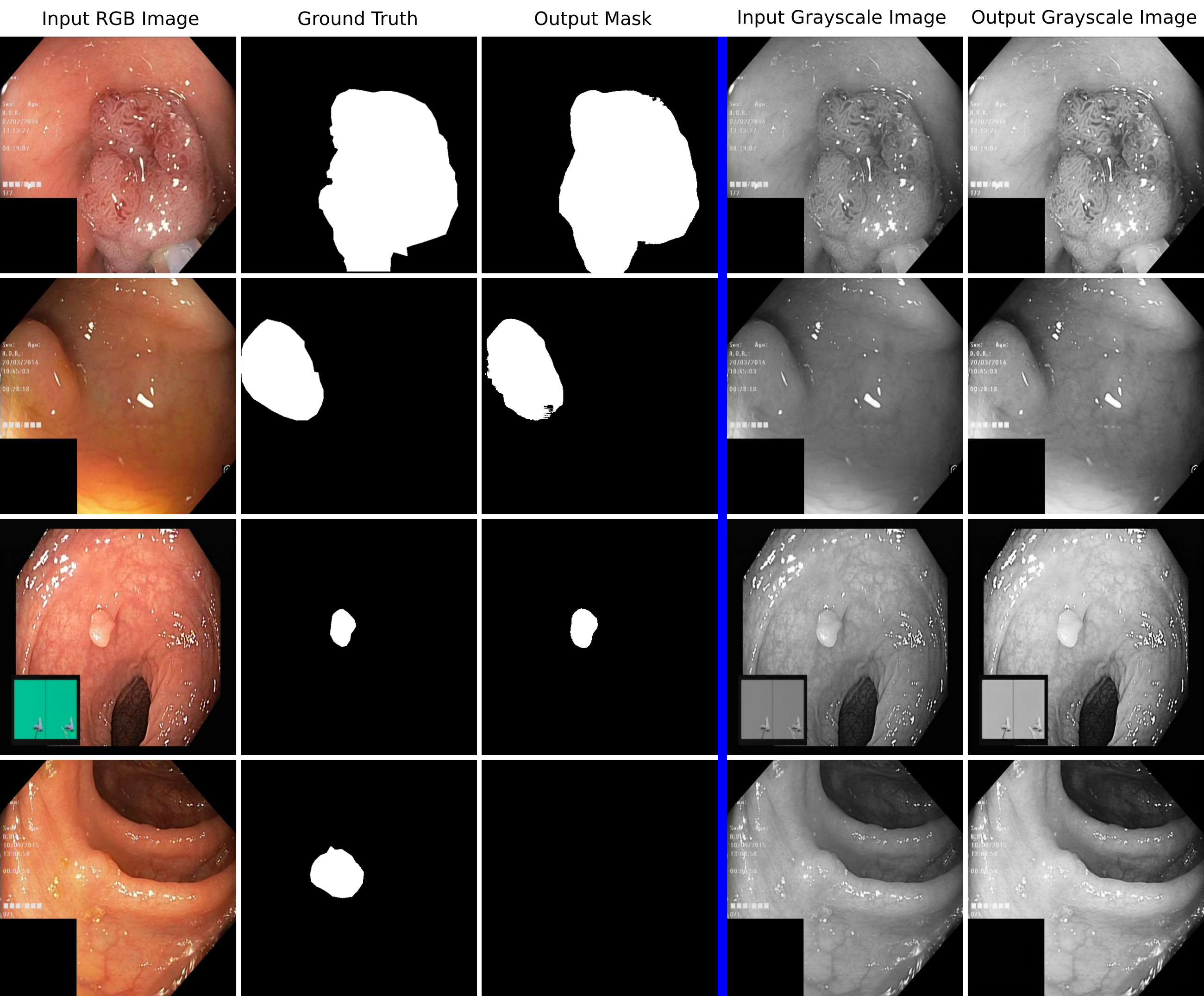}
    \caption{Qualitative results of the DDANet on the Kvasir-SEG test dataset. The blue line divides the segmentation and the reconstruction part. Columns 4 and 5 show the reconstruction part that was used in the DDANet as an auxiliary task.}
    \label{fig:kvasirsegqualitative}
    \vspace{-5mm}
\end{figure}

Table~\ref{table:resultkvasir} shows the results of the DDANet trained and validated on Kvasir-SEG. Additionally, evaluation scores on the test dataset can also be found here. The evaluation metrics for the challenge was \ac{DSC}. However, we have also calculated other commonly used metrics such as \ac{mIoU}, recall, precision, and \ac{FPS}. The DDANet obtained a \ac{DSC} of 0.8576, a \ac{mIoU} of 0.7800, a recall of 0.8880, and a precision of 0.8643. All the metrics suggest that our method performs quite well on the Kvasir-SEG dataset. When we compare the results with our previous results~\cite{jha2019resunet++,jha2020kvasir}, where the \ac{DSC} values were 0.8133, and 0.7877, DDANet achieves a higher \ac{DSC} of 0.8576. However, we can not compare directly with this work with our previous works as a different train-test split of the dataset is used.  

Figure~\ref{fig:kvasirsegqualitative} shows the qualitative results of the DDANet on Kvasir-SEG. The figure shows that the proposed DDANet is able to segment both larger and smaller polyps. However, the figure also shows the challenges in identifying the flat polyps, which is one of the open issues in the field of development of \ac{CADx} systems for colonoscopy. From the quantitative results on development and unseen test dataset, we can say that the proposed method is comprehensive in producing reliable segmentation output.

\section{Discussion}
\label{discussion}
\vspace{-3mm}
The qualitative results (see Figure~\ref{fig:kvasirsegqualitative}) show that the proposed model was able to segment polyps ranging from large to small (Figure~\ref{fig:kvasirsegqualitative}), but still, challenges remain within some polyps (for example, flat or sessile). We can also see a nearly perfect reconstruction of the grayscale image. In the future, we would like to use image super-resolution instead of just a grayscale image reconstruction.   

From all the results, we can see that our method achieves high precision and recall evaluation scores on both the Kvasir-SEG validation dataset and on the unseen test dataset (see Table~\ref{table:resultkvasir}). Additionally, we also achieved a DSC of  0.7874 on the unseen dataset. Thus, high DSC, recall, and precision results validate our proposed method.  Moreover, our approach is quite fast with an average \ac{FPS} of 70.23. Thus, the results show that our method can identify polyps in real-time.    

\section{Conclusion}
\label{sec:Conclusion}
\vspace{-3mm}
The Endotect challenge~\cite{Hicks2020} aims to benchmark various computer-vision approaches on the HyperKvasir dataset containing \ac{GI} images and videos. Here, we have proposed the \ac{DDANet} architecture for automatic polyp segmentation, and the proposed architecture provides good results in the segmentation task. We have obtained a high precision, recall, \ac{DSC}, \ac{mIoU}, and \ac{FPS}. However, there are large rooms for improvements. We intend to further improve the architecture by applying post-processing and analyzing the optimal parameters in the future. 

\vspace{-3mm}
\section*{Acknowledgements}
\vspace{-3mm}
This work is funded in part by the Research Council of Norway, project number 263248 (Privaton) and project number 282315 (AutoCap). We performed all computations in this paper on equipment provided by the Experimental Infrastructure for Exploration of Exascale Computing ($eX^3$), which is financially supported by the Research Council of Norway under contract 270053.
\bibliographystyle{splncsnat}
\vspace{-3mm}
\bibliography{references.bib}
\end{document}